\documentclass[12pt,draftcls,onecolumn]{IEEEtran}

\usepackage{etoolbox}
\makeatletter
\def\ps@headings{%
	\def\@oddhead{\mbox{}\scriptsize\rightmark \hfil \thepage}%
	\def\@evenhead{\scriptsize\thepage \hfil \leftmark\mbox{}}%
	\def\@oddfoot{}%
	\def\@evenfoot{}}
\makeatother
\usepackage{verbatim}
\usepackage{xcolor}
\usepackage{amsfonts}
\usepackage{mathrsfs}
\usepackage{amsfonts}
\usepackage{amssymb}
\usepackage{graphicx}
\usepackage{epsfig}
\usepackage{epstopdf}
\usepackage{psfrag}
\usepackage{amsmath}
\usepackage{array}
\usepackage{cases}
\usepackage{subfigure}
\usepackage{cite,graphicx,amsmath,amssymb,color}
\usepackage{algorithmic}
\usepackage{algorithm}

\usepackage{algorithm}
\usepackage{algorithmic}
\usepackage{stmaryrd}
\usepackage{multirow}
\usepackage{subfig}
\usepackage{graphicx,times,amsmath} 
\DeclareMathOperator*{\argmax}{argmax}

\usepackage{url}
\usepackage{enumerate}

\IEEEoverridecommandlockouts

\newtheorem{lemma}{Lemma}

\newtheorem{Exam}{Example}

\newtheorem{problem}{Problem}
\newtheorem{remark}{Remark}

\newtheorem{case}{Case}

\begin{document}

\title{Energy-Efficient Multi-View Video Transmission with View Synthesis-Enabled Multicast}

\author{\authorblockN{Wei Xu, \quad Yuzhuo Wei, \quad Ying Cui, \quad Zhi Liu}\thanks{W. Xu, Y. Wei and Y. Cui  are with Shanghai Jiao Tong University, China. Z. Liu is with Shizuoka  University, Japan.}}


\maketitle

\begin{abstract}
Multi-view videos (MVVs) provide  immersive viewing experience, at the cost of heavy load to wireless networks. Except for further improving viewing experience, view synthesis can create multicast opportunities for efficient transmission of MVVs in multiuser wireless networks, which has not been recognized in existing literature. In this paper, we would like to exploit view synthesis-enabled multicast opportunities for energy-efficient MVV transmission in a  multiuser wireless network.
Specifically, we first establish a mathematical model to characterize the impact of view synthesis on multicast opportunities and energy consumption. Then, we consider the optimization of view selection, transmission time and power allocation to minimize the weighted sum energy consumption for view transmission and synthesis, which is a challenging mixed discrete-continuous optimization problem. We propose an algorithm to obtain an optimal solution with reduced computational complexity by exploiting optimality properties. To further reduce computational complexity, we also propose two low-complexity algorithms to obtain two suboptimal solutions, based on continuous relaxation and Difference of Convex (DC) programming, respectively. Finally, numerical results demonstrate the advantage of the proposed solutions.
\end{abstract}



%
\section{Introduction}
A multi-view video (MVV) is generated by capturing a scene of interest with multiple cameras from different angles simultaneously. Each camera can capture both texture maps (i.e., images) and depth maps (i.e., distances from objects in the scene), providing one view. Besides views captured by cameras, additional views, referred to as virtual views, can be synthesized based on reference views, providing new view angles to further enhance viewing experience. More specifically, each virtual view can be synthesized based on a left view and a right view using Depth-Image-Based Rendering (DIBR) \cite{fehn2004depth}. A MVV subscriber (i.e., user) can freely select among multiple view angles, hence enjoying immersive viewing experience. MVV has vast applications in entertainment, education, medicine, etc. For example, MVV is one key technique in free-viewpoint television, naked-eye 3D and virtual reality (VR).

A MVV is in general of a much larger size than a traditional single-view video, bringing a heavy burden to wireless networks. The coding structure of a MVV (i.e. how the MVV frames are arranged and encoded) determines the traffic load on a wireless network. To facilitate MVV transmission, views are usually encoded separately using standard video codec and only the view corresponding to a user's current selected viewpoint is transmitted \cite{liu2013optimizing,toni2017optimal,toni2016network,zhang2017optimized}.

In \cite{toni2017optimal,toni2016network,zhang2017optimized}, the authors consider a wired MVV system with a single server and multiple users. Note that view synthesis usually introduces distortion, the degree of which depends on the distance between the two reference views and their qualities. References \cite{toni2017optimal,toni2016network,zhang2017optimized} consider the optimization of view selection to minimize the total distortion of all synthesized views subject to the bandwidth constraint. As the transmission models in \cite{toni2017optimal,toni2016network,zhang2017optimized} do not reflect channel fading and broadcast nature which are key features of wireless networks, the solutions of MVV transmission in \cite{toni2017optimal,toni2016network,zhang2017optimized} cannot be applied to MVV transmission in wireless networks.

In \cite{zhao2015qos} and \cite{zhao2014qos}, the authors consider a wireless MVV transmission system with a single server and multiple users, where channel fading and broadcast nature of wireless communications are captured. The transmission mechanisms in \cite{zhao2015qos} and \cite{zhao2014qos} make use of natural multicast opportunities to reduce energy consumption. In particular, \cite{zhao2014qos} considers Orthogonal Frequency Division Multiple Access (OFDMA), and optimizes power and subcarrier allocation to minimize the total transmission power. Neither of \cite{zhao2015qos} and \cite{zhao2014qos} considers view synthesis at the server or users, which can create multicast opportunities to further improve transmission efficiency and reduce energy consumption in wireless networks. As far as we know, this benefit of view synthesis has not been recognized in existing literature. Thus, the performance of the transmission designs in \cite{zhao2015qos} and \cite{zhao2014qos} may be further improved.

In this paper, we would like to address the above limitation. We consider MVV transmission from a server to multiple users in a wireless network using Time Division Multiple Access (TDMA). Different from \cite{zhao2015qos} and \cite{zhao2014qos}, we allow view synthesis at the server and each user to create multicast opportunities for efficient MVV transmission in multiuser wireless networks. Specifically, we first establish a mathematical model to characterize the impact of view synthesis on multicast opportunities and energy consumption. Then, we consider the optimization of view selection, transmission time and power allocation to minimize the weighted sum energy consumption for view transmission and synthesis. The problem is a challenging mixed discrete-continuous optimization problem. We propose an algorithm to obtain an optimal solution with reduced computational complexity by exploiting optimality properties. To further reduce computational complexity, we propose two low-complexity algorithms to obtain two suboptimal solutions. Specifically, the first suboptimal solution is obtained by transforming the continuous relaxation of the original problem into a convex problem and rounding the optimal solution of the convex problem. The second suboptimal solution is obtained by transforming the original problem into a Difference of Convex (DC) programming problem and finding a stationary point using a DC algorithm \cite{lipp2016variations}. The second suboptimal solution achieves lower energy consumption with higher computational complexity than the first suboptimal solution. To the best of our knowledge, this is the first work providing optimization-based solutions for energy-efficient MVV transmission by effectively exploiting view synthesis-enabled multicast opportunities in multiuser wireless networks. Finally, numerical results demonstrate the advantage of  the proposed suboptimal solutions.
\section{System Model}
\begin{figure}[!t]
	\centering
	\includegraphics[width=9cm,height=4cm]{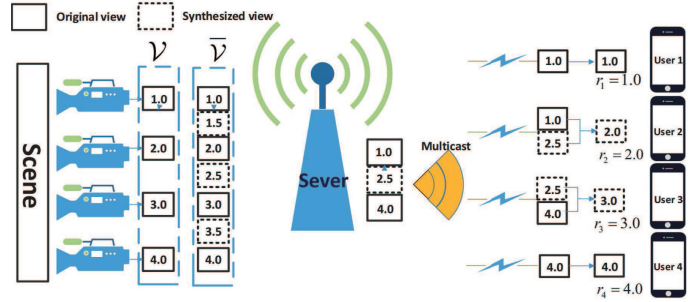}
	\caption{System model. $K=4$, $r_1=1$, $r_2=2$, $r_3=3$, $r_4=4$, $\mathcal{V}=\{1,2,3,4\}$, $\overline{\mathcal{V}}=\{1,1.5,2,\cdots,4\}$, $\Delta=1$, $x_1=x_{2.5}=x_{4}=1$ and $y_{1,1}=y_{2,1}=y_{2,2.5}=y_{3,2.5}=y_{3,4}=y_{4,4}=1$.}
	\label{system model}
\end{figure}

As illustrated in Fig.~\ref{system model}, we consider downlink transmission of a MVV from a single-antenna server (e.g., base station or access point) to $K$ $(>1)$ single-antenna users. $V$ views (including texture maps and depth maps) about a scene of interest are captured by $V$ evenly spaced cameras simultaneously from different view angles. The $V$ views are then pre-encoded independently using standard video codec and stored at the server. Let $\mathcal{V}=\{1,2,\cdots,V\}$ denote the set of indices for the original views. We consider $Q-1$ evenly spaced additional views between view $v$ and $v+1$ with view spacing $1/Q$ between two neighboring views, where $Q=2,3,\cdots$ is a system parameter and $v\in\{1,2,\cdots,V-1\}$. The additional views can be synthesized via DIBR to provide new view angles. The set of indices for all views including the $V$ original views (which are stored at the server) and the $(V-1)(Q-1)$ additional views (which are not stored at the server but can be synthesized on demand) is denoted by $\overline{\mathcal{V}}=\{1,1+1/Q,1+2/Q,\cdots,V\}$. For ease of exposition, we assume all views have the same source encoding rate (in bits/s), denoted by $R$.

Using  DIBR, a view can be synthesized by using one left view and one right view as the reference views, either by the server or by a user. The quality of each synthesized view depends on its distance to its two reference views and the qualities of its reference views. The server only needs to synthesize additional views. Specifically, it can synthesize each additional view $v\in \overline{\mathcal{V}} \setminus \mathcal{V}$ using its nearest left original view $\lfloor v \rfloor$ and right original view $\lceil v \rceil$.\footnote{$\lfloor v \rfloor$ denotes the greatest integer less than or equal to $v$ and $\lceil v \rceil$ denotes the least integer greater than or equal to $v$.} Each user may need to synthesize any view $v\in \overline{\mathcal{V}}$, by using two views from the left reference view set $\overline{\mathcal{V}}_{v}^- = \{x\in \overline{\mathcal{V}}: v-\Delta\leq x < v\}$ and the right reference view set $\overline{\mathcal{V}}_{v}^+=\{x\in\overline{\mathcal{V}}:v<x\leq v+\Delta\}$, respectively. Note that $\overline{\mathcal{V}}_1^- = \emptyset$ and $\overline{\mathcal{V}}_{V}^+=\emptyset$. Here, $\Delta~(\geq 1)$ is a system parameter to limit the distance between each synthesized view and each of its reference views so as to guarantee the quality of each synthesized view.

Let $\mathcal{K} =\{1,\ldots,K\}$ denote the set of user indices. Let $r_k \in \overline{\mathcal{V}}$ denote the index of the view requested by user $k\in \mathcal{K}$. Note that different users can request the same view, corresponding to natural multicast opportunities. To satisfy user $k$'s view request, the server either transmits view $r_k$ or transmits two reference views in $\overline{\mathcal{V}}_{r_k}^-$ and $\overline{\mathcal{V}}_{r_k}^+$ for user $k$ to synthesize view $r_k$. To save resource, the server transmits each view at most once, making use of both natural multicast opportunities and view synthesis-enabled multicast opportunities (which will be further illustrated in Example~1). Let $x_v$ denote the view transmission variable for view $v$, where $x_v$ satisfies:
\begin{equation}
x_v\in\{0,1\},\quad v\in \overline{\mathcal{V}}. \label{binary constraint x}
\end{equation}
Here, $x_v=1$ indicates that the server will transmit view $v$ and $x_v=0$ otherwise. Denote $\mathbf{x} \triangleq (x_v)_{v \in \overline{\mathcal{V}}}$. Let $y_{k,v}$ denote the view utilization variable for view $v$ at user $k$, where $y_{k,v}$ satisfies:
\begin{equation}
y_{k,v}\in\{0,1\},\quad v\in \overline{\mathcal{V}},\ k \in \mathcal{K}. \label{binary constraint y}
\end{equation}
Here, $y_{k,v}=1$ indicates that user $k$ will utilize view $v$ (as view $v$ is requested by user $k$, i.e., $r_k=v$, or view $v\in \overline{\mathcal{V}}_{r_k}^+$ or $\overline{\mathcal{V}}_{r_k}^-
$ is used to synthesize view $r_k$ at user $k$) and $y_{k,v}=0$ otherwise. Denote $\mathbf{y} \triangleq \left( y_{k,v} \right)_{k\in \mathcal{K} ,v\in \overline{\mathcal{V}}}$. Thus, to guarantee that each user can obtain its requested view, we require:
\begin{align}
&y_{k,r_k}+\sum_{v\in \overline{\mathcal{V}}_{r_k}^{+}} y_{k,v} = 1, \quad k\in \mathcal{K},  \label{Right constraint} \\
&y_{k,r_k}+\sum_{v\in \overline{\mathcal{V}}_{r_k}^{-}} y_{k,v} = 1, \quad k\in \mathcal{K}. \label{Left constraint}
\end{align}
Note that the constraints in (\ref{binary constraint y}), (\ref{Right constraint}) and (\ref{Left constraint}) ensure that either $y_{k,r_k}=1$, $\sum_{v\in \overline{\mathcal{V}}_{r_k}^{+}} y_{k,v}=\sum_{v\in \overline{\mathcal{V}}_{r_k}^{-}} y_{k,v}=0$, or $y_{k,r_k}=0$, $\sum_{v\in \overline{\mathcal{V}}_{r_k}^{+}} y_{k,v}=\sum_{v\in \overline{\mathcal{V}}_{r_k}^{-}} y_{k,v}=1$. The server has to transmit view $v$ in order for a user to utilize view $v$. Thus, we have the following constraint connecting the view transmission variables and view utilization variables:
\begin{equation}
	x_v \geq y_{k,v}, \quad k\in \mathcal{K},\ v\in \overline{\mathcal{V}}. \label{x>y}
\end{equation}
View selection, reflecting view synthesis, is achieved by choosing $\mathbf{x}$ and $\mathbf{y}$.
\begin{Exam}\textit{(View Synthesis-Enabled Multicast Opportunities):}
	Consider an illustration example as shown in Fig 1. In this example, without view synthesis, the server has to transmit four views, i.e., views 1, 2, 3 and 4, and there are no natural multicast opportunities (as different users request different views). In contrast, if view synthesis is allowed at the server and each user, the server can transmit only three views, i.e., views 1, 2.5 and 4, and each of the three views can be utilized by two users. Therefore, view synthesis can create multicast opportunities, enabling more efficient transmission designs for MVVs in multiuser wireless networks.
\end{Exam}

We consider Time Division Multiple Access (TDMA). Each TDMA frame is of duration $T$ (in seconds). Consider one frame. The time allocated to transmit view $v$, denoted by $t_v$, satisfies
\begin{equation}
t_v \geq 0,\quad  v\in \overline{\mathcal{V}}. \label{t>=0}
\end{equation}
In addition, we have the following total time allocation constraint:
\begin{equation}
\sum_{v\in \overline{\mathcal{V}}} t_v \leq T. \label{time constraint}
\end{equation}

We consider a narrow band system and let $B$ denote the bandwidth (in Hz). We study the block fading channel model. Let $h_k$ denote the channel power for user $k$, which is assumed to be constant within each TDMA frame. Different views are encoded separately and transmitted over different time. Let $p_v$ denote the transmission power for view $v$. Then, the maximum transmission rate (in bits/s) of view $v$ to user $k$ is given by $B\log_2\left(1+\frac{p_v h_k}{n_0} \right)$, where $n_0$ is the power of the complex additive white Gaussian channel noise at each receiver. To guarantee that all users that need to utilize view $v$ can successfully decode it, we have the following successful transmission constraint:
\begin{equation}
t_vB\log_2\left(1+\frac{p_v h_k}{n_0}\right) \geq  y_{k,v} RT, \quad k\in \mathcal{K},\  v\in \overline{\mathcal{V}}. \label{bandwidth constraint}
\end{equation}
In other words, we consider multicast transmission of view $v$, if there are more than one user utilizing it. The transmission energy consumption at the server is given by:
\begin{equation}
E_\text{t}(\mathbf{t},\mathbf{p})=\sum_{v\in \overline{\mathcal{V}}} t_vp_v,
\end{equation}
where $\mathbf{t}\triangleq (t_v)_{v\in \overline{\mathcal{V}}}$ and $\mathbf{p}\triangleq (p_v)_{v\in \overline{\mathcal{V}}}$. Let $E_b$ denote the synthesis energy consumption (caused by computation) \cite{6195536} at the server for one view. Then, the total synthesis energy consumption at the server is given by:
\begin{equation}
E_{\text{s}}^{(b)}(\mathbf{x})=\sum_{v\in \overline{\mathcal{V}} \setminus \mathcal{V}}x_vE_b,
\end{equation}
where $\mathbf{x}\triangleq (x_v)_{v\in \overline{\mathcal{V}}}$.
Similarly, let $E_{\text{u},k}$ denote the synthesis energy consumption at user $k$ for one view. Note that we allow $E_{\text{u},k},k\in \mathcal{K}$ to be different due to heterogenous hardware conditions at different users. Then, the total synthesis energy consumption at all users is given by:
\begin{equation}
E_{\text{s}}^{(u)}(\mathbf{y})=\sum_{k\in \mathcal{K}} (1-y_{k,r_k})E_{\text{u},k}.
\end{equation}
Therefore, the weighted sum energy consumption is given by:
\begin{equation}
E(\mathbf{t},\mathbf{p},\mathbf{x},\mathbf{y}) = E_\text{t}(\mathbf{t},\mathbf{p})+E_{\text{s}}^{(b)}(\mathbf{x})+\beta E_{\text{s}}^{(u)}(\mathbf{y}),
\end{equation}
where $\beta\geq 1$ is the corresponding weight factor. Note that $\beta >1 $ means imposing a higher cost on energy consumption for user devices due to their limited battery power.
\begin{remark}\textit{(Modeling of View Synthesis and Multicast in Multiuser Wireless Networks):}
	The proposed model based on view transmission variables $\mathbf{x}$ and view utilization variables $\mathbf{y}$ mathematically characterizes the impact of view synthesis on multicast opportunities and energy consumption at the server and each user. Later, we shall see that this enables optimizing view synthesis-based multicast opportunities for transmission energy reduction.
\end{remark}
\section{Problem Formulation and Optimal Solution}
\subsection{Problem Formulation}
We would like to minimize the weighted sum energy consumption by optimizing the view transmission and utilization variables (i.e., view selection) as well as the transmission time and power allocation variables. Specifically, we have the following optimization problem.
\begin{problem}[Energy Minimization]\label{View selection and resource allocation}
\begin{align}
	E^\star \triangleq &\min_{\mathbf{x},\mathbf{y},\mathbf{p},\mathbf{t}}\quad E(\mathbf{t},\mathbf{p},\mathbf{x},\mathbf{y}) \notag\\
	&~~\text{s.t.} \quad (\ref{binary constraint x}),(\ref{binary constraint y}),(\ref{Right constraint}),(\ref{Left constraint}),(\ref{x>y}),(\ref{t>=0}),(\ref{time constraint}),(\ref{bandwidth constraint}). \notag
\end{align}
\end{problem}
Let ($\mathbf{x}^\star,\mathbf{y}^\star,\mathbf{p}^\star,\mathbf{t}^\star$) denote the optimal solution of Problem \ref{View selection and resource allocation}.

Obviously, Problem \ref{View selection and resource allocation} is a mixed discrete-continuous optimization problem with two types of variables, i.e., view transmission and utilization variables (binary variables) as well as power allocation and time allocation variables (continuous variables). In general, Problem~\ref{View selection and resource allocation} is NP-hard.
\begin{figure}[!t]
	\centering
	\includegraphics[width=\linewidth,height=6cm]{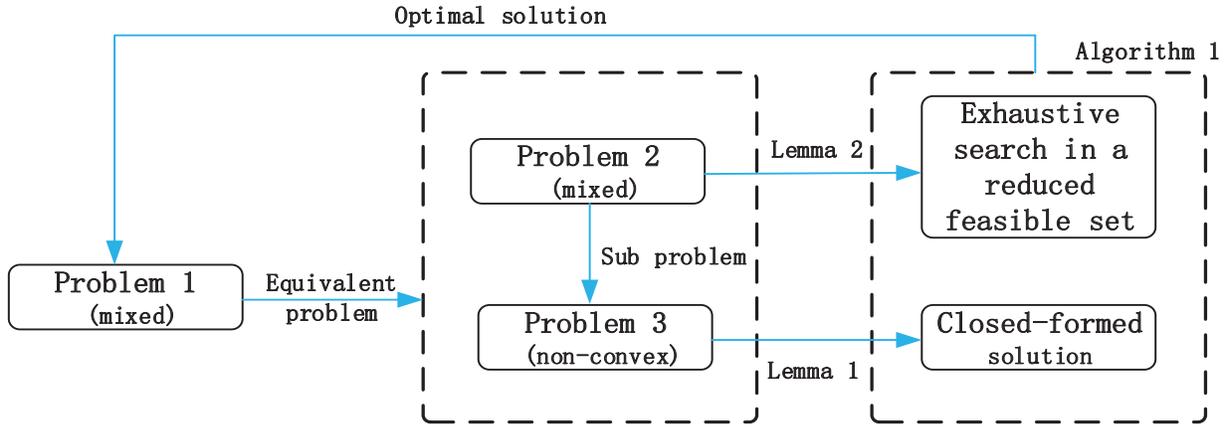}
	\caption{\small{Proposed optimal solution of Problem 1.}}
	\label{structure_1}
\end{figure}
\subsection{Optimal Solution}
In this section, we develop an algorithm to obtain an optimal solution of Problem~\ref{View selection and resource allocation}, as shown in Fig.~\ref{structure_1}. Define $\mathbf{Y} \triangleq \{ \mathbf{y}: (\ref{binary constraint y}),(\ref{Right constraint}),(\ref{Left constraint}) \}$ and $\mathbf{X}\times \mathbf{Y} \triangleq \{(\mathbf{x},\mathbf{y}): (\ref{binary constraint x}),(\ref{x>y}), \mathbf{y}\in \mathbf{Y}\}$. First, by exploiting structural properties of Problem~\ref{View selection and resource allocation}, we propose an equivalent formulation of Problem~\ref{View selection and resource allocation}.
\begin{problem}[View Transmission and Utilization]\label{View selection}
\begin{align}
    &\min_{(\mathbf{x},\mathbf{y}) \in \mathbf{X}\times \mathbf{Y}} E_\text{t}^\star(\mathbf{x},\mathbf{y})+E_\text{s}^{(b)}(\mathbf{x})+\beta E_\text{s}^{(u)}(\mathbf{y}) \notag
\end{align}
\end{problem}
where $E_\text{t}^\star(\mathbf{x},\mathbf{y})$ is given by the following sub-problem. Let $\mathbf{x}^\star \triangleq (x_v^\star)_{v\in \overline{\mathcal{V}}}$ and $\mathbf{y}^\star \triangleq (y^\star_{k,v})_{k\in \mathcal{K},v\in \overline{\mathcal{V}}}$ denote the optimal solution of Problem~\ref{View selection}.
\begin{problem}[Time and Power Allocation for Given $\mathbf{x}$,$\mathbf{y}$]\label{sub-problem for resource allocation} For given $\mathbf{x}$ and $\mathbf{y}$ that satisfy (\ref{binary constraint x})-(\ref{x>y}), we have:
	\begin{align}
	E_\text{t}^\star(\mathbf{x},\mathbf{y}) \triangleq&\min_{\mathbf{t},\mathbf{p}} \quad E_\text{t}(\mathbf{t},\mathbf{p}) \notag\\
	&~\text{s.t.} \quad (\ref{t>=0}),(\ref{time constraint}),(\ref{bandwidth constraint}). \notag
	\end{align}
\end{problem}
Let $\mathbf{t}^\star(\mathbf{x},\mathbf{y})\triangleq (t^\star_v \left(\mathbf{x},\mathbf{y})\right)_{v\in\overline{\mathcal{V}}}$ and $\mathbf{p}^\star(\mathbf{x},\mathbf{y})\triangleq (p^\star_v \left(\mathbf{x},\mathbf{y})\right)_{v\in\overline{\mathcal{V}}}$ denote the optimal solution of Problem~\ref{sub-problem for resource allocation}.

This formulation (including Problem~\ref{View selection} and Problem~\ref{sub-problem for resource allocation}) separates the two types of variables (i.e., binary variables and continuous variables) and facilitates the optimization. We can obtain an optimal solution of Problem~\ref{View selection and resource allocation} by solving Problem~\ref{View selection} and Problem~\ref{sub-problem for resource allocation}. First, we focus on solving Problem~\ref{sub-problem for resource allocation}. As $E_\text{t}(\mathbf{t},\mathbf{p})$ is not convex in $(\mathbf{t},\mathbf{p})$, Problem~\ref{sub-problem for resource allocation} is nonconvex. By exploiting optimality properties of Problem~\ref{sub-problem for resource allocation}, we can obtain the closed-form optimal solution.
\begin{lemma}[Optimal Solution of Problem~\ref{sub-problem for resource allocation}]\label{subproblem}
An optimal solution of Problem~\ref{sub-problem for resource allocation} is given by:
\begin{align}
&t_v^\star(\mathbf{x},\mathbf{y})= \begin{cases}
\frac{RT \log2}{B\left( W(\frac{\lambda^\star(\mathbf{x},\mathbf{y}) h_{v,\min}(\mathbf{y}_v)}{n_0 e}-\frac{1}{e})+1\right)}, & x_v=1\\
0, & x_v=0 \end{cases}
,\ v \in \overline{\mathcal{V}}, \notag \\
&p_v^\star(\mathbf{x},\mathbf{y})=\begin{cases}
\frac{n_0}{h_{v,\min}(\mathbf{y}_v)}\left( 2^{\frac{RT}{Bt_v^\star(\mathbf{x},\mathbf{y})}}-1\right), & x_v=1  \notag\\
0, & x_v=0
\end{cases}
,\ v \in \overline{\mathcal{V}},
\end{align}
where $\mathbf{y}_v \triangleq (y_{k,v})_{k\in \mathcal{K}}$, and $\lambda^\star(\mathbf{x},\mathbf{y})$ satisfies
\begin{equation}
\sum_{v\in \overline{\mathcal{V}}} \frac{RT\log 2}{B\left( W(\frac{\lambda^\star(\mathbf{x},\mathbf{y}) h_{v,\min}(\mathbf{y}_v)}{n_0 e}-\frac{1}{e})+1\right)}x_v=T. \notag
\end{equation}
Here, $W(\cdot)$ denotes lambert W function and
$h_{v,\min}(\mathbf{y}_v)\triangleq \min_{k\in \mathcal{K}} h_ky_{k,v}$ for all $v \in \overline{\mathcal{V}}$.
\begin{proof}
	Please refer to Appendix A.
\end{proof}
\end{lemma}

\begin{figure}[!ht]
	\centering
	\includegraphics[width=9cm,height=7cm]{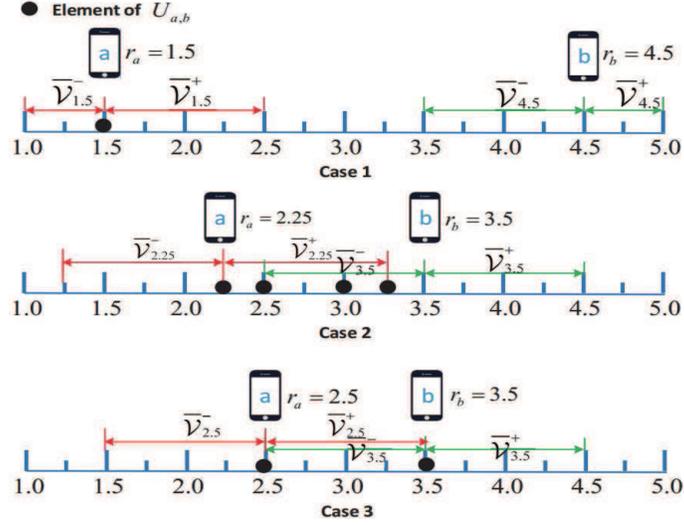}
 	\caption{Illustration of $U_{a,b}$. $\mathcal{V}=\{1,2,3,4,5\}$, $\overline{\mathcal{V}}=\{1,1.25,1.5,\cdots,5\}$ and $\Delta=1$.}
	\label{Illustration}
\end{figure}

Next, we focus on solving Problem~\ref{View selection}. Problem~\ref{View selection} is a discrete optimization problem and is NP-hard. Problem~\ref{View selection} can be solved by exhaustive search. To reduce the search space, we first analyze optimality properties of Problem~\ref{View selection}. Consider any two users $a \in \mathcal{K}$ and $b \in \mathcal{K}$ and define $r_{\max} \triangleq \max \{r_a,r_b\}$ and $r_{\min} \triangleq \min \{r_a,r_b\}$. Study three cases, i.e., Case 1: $\overline{\mathcal{V}}_{r_{\min}}^+ \cap \overline{\mathcal{V}}_{r_{\max}}^- = \emptyset$, Case 2: $\overline{\mathcal{V}}_{r_{\min}}^+ \cap \overline{\mathcal{V}}_{r_{\max}}^- \neq \emptyset$ and $r_{\max} \notin \overline{\mathcal{V}}_{r_{\min}}^+$, and Case 3: $r_{\max} \in \overline{\mathcal{V}}_{r_{\min}}^+$. We define:
\begin{align}
&U_{a,b} \triangleq \notag \begin{cases}
\{r_a\}, & \text{Case~1}\\
\{ r_a,r_{\max}-\Delta,r_{\min}+\Delta\}\cup (\overline{\mathcal{V}}_{r_{\min}}^+ \cap \overline{\mathcal{V}}_{r_{\max}}^- \cap \mathcal{V}), & \text{Case~2} \\
\{ r_a,r_b,r_{\max}-\Delta,r_{\min}+\Delta\},& \text{Case~3},
\end{cases} \notag \\
&U_k \triangleq \bigcup_{i \in \mathcal{K}:i\neq k} U_{k,i}. \notag
\end{align}
Note that $U_{a,b}$ characterizes the set of views that may be utilized by user $a$ when considering only users $a$ and $b$, as illustrated in Fig.~\ref{Illustration}, and $U_k$ specifies the set of views that may be utilized by user $k$ considering all users. Then, we have the following lemma.
\begin{lemma}[Optimality Properties of Problem~\ref{View selection}] \label{optimality properties}
(i) $x_v^\star = \max_{k\in \mathcal{K}} y_{k,v}^\star~ \text{for all } v \in \overline{\mathcal{V}}$;
(ii) Suppose $\beta E_{\text{u},k}\geq E_{\text{b}}$ for all $k \in \mathcal{K}$, if $ v \in \overline{\mathcal{V}}$ and $v \notin U_k$ for all $k \in \mathcal{K}$, then $y^\star_{k,v}=0$ for all $k\in \mathcal{K}$.
\begin{proof}
	Please refer to Appendix B.
\end{proof}
\end{lemma}
\
Statement (i) indicates that view $v$ will not be transmitted if no user utilizes it. Statement 
(ii) indicates that any view not in $U_k$ will not be utilized by user $k$. Let $\underline{\mathbf{X}} \times \underline{\mathbf{Y}} \triangleq \{ (\mathbf{x},\mathbf{y}) |\ \mathbf{y}\in \underline{\mathbf{Y}},x_v= \max_{k\in\mathcal{K}} y_{k,v} \text{ for all } v\in \overline{\mathcal{V}}  \}$, where $\underline{\mathbf{Y}} \triangleq \{\mathbf{y} \in \mathbf{Y}|y_{k,v}=0 \text{ for all } v \notin U_k, k\in \mathcal{K}\}$. Based on Lemma~\ref{optimality properties}, we can reduce the feasible set for $(x,y)$ from $\mathbf{X} \times \mathbf{Y}$ to $\underline{\mathbf{X}} \times \underline{\mathbf{Y}}$ without loss of optimality. Therefore, by Lemma~\ref{subproblem} and Lemma~\ref{optimality properties}, we develop an algorithm to obtain an optimal solution of Problem~\ref{View selection and resource allocation}, as summarized in Algorithm~1.
\begin{algorithm}[!t]
	\caption{Optimal solution of Problem~\ref{View selection and resource allocation}}
	\textbf{Output} ($\mathbf{x}^\star$,$\mathbf{y}^\star$,$\mathbf{p}^\star$,$\mathbf{t}^\star$).
	\begin{algorithmic}[1]
	\STATE Set $E^\star=\infty$.
	\FOR{$(\mathbf{x},\mathbf{y}) \in \underline{\mathbf{X}} \times \underline{\mathbf{Y}}$}
	\STATE For given $(\mathbf{x},\mathbf{y})$, obtain $(\mathbf{p},\mathbf{t})$ based on Lemma~\ref{subproblem} and compute $E(\mathbf{t},\mathbf{p},\mathbf{x},\mathbf{y})=E_\text{t}^\star(\mathbf{x},\mathbf{y})+E_\text{s}^{(b)}(\mathbf{x})+\beta E_\text{s}^{(u)}(\mathbf{y})$.
	\IF{$E(\mathbf{t},\mathbf{p},\mathbf{x},\mathbf{y}) \leq E^\star$}
	\STATE Set $E^\star = E(\mathbf{t},\mathbf{p},\mathbf{x},\mathbf{y})$ and  $(\mathbf{x}^\star$,$\mathbf{y}^\star$,$\mathbf{p}^\star$,$\mathbf{t}^\star) =( \mathbf{x}$,$\mathbf{y}$,$\mathbf{p}$,$\mathbf{t})$.
	\ENDIF
	\ENDFOR
	\end{algorithmic}
\end{algorithm}

\section{Suboptimal Solutions}
\begin{figure*}[!t]
	\centering
	{\resizebox{\linewidth}{!}{\includegraphics[width=\linewidth,height=7cm]{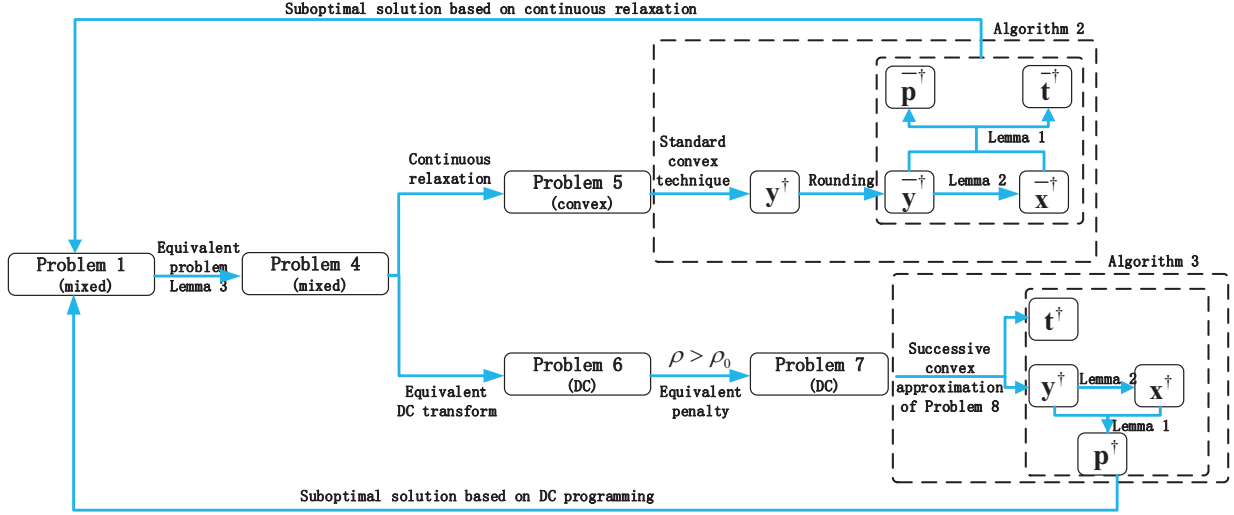}}}
	\caption{\small{Proposed suboptimal solutions of Problem 1.}}
	\label{structure_2}
\end{figure*}
Although the complexity for obtaining an optimal solution of Problem~\ref{View selection} has been reduced based on Lemma~\ref{optimality properties}, the complexity of Algorithm 1 is still unacceptable when $K$ is large. In this section, we propose two low-complexity algorithms to obtain suboptimal solutions of Problem~\ref{View selection and resource allocation}, as illustrated in Fig.~\ref{structure_2}. First, define:
\begin{align}
&E_1(\mathbf{t},\mathbf{y})\triangleq\sum_{v\in \overline{\mathcal{V}}} t_v \max_{k\in \mathcal{K}} \left\{ \frac{n_0}{h_k} \left(2^{\frac{y_{k,v}RT}{Bt_v}}-1\right)\right\},\\
&E_2(\mathbf{y})\triangleq E_b\sum_{v\in \overline{\mathcal{V}} \setminus \mathcal{V}}\max_{k \in \mathcal{K}} y_{k,v},\\
&E_3(\mathbf{y})\triangleq\sum_{k \in \mathcal{K}} E_{u,k} \sum_{v\in \overline{\mathcal{V}}_{r_k}^+}y_{k,v}.
\end{align}
Next, introduce the following problem.
\begin{problem}[View Utilization and Transmission Time]	\label{view selection and time allocation}
	\begin{align}
	E^*\triangleq&\min_{\mathbf{t},\mathbf{y}}~ E_1(\mathbf{t},\mathbf{y})+E_2(\mathbf{y})+E_3(\mathbf{y}) \notag\\
	&\text{ s.t.} \quad  (\ref{binary constraint y}),(\ref{Right constraint}),(\ref{Left constraint}),(\ref{t>=0}),(\ref{time constraint}). \notag
	\end{align}
\end{problem}
Let $\mathbf{t}^*$ and $\mathbf{y}^*$ denote the optimal solution of Problem~\ref{view selection and time allocation}.

By exploiting optimality properties of Problem~\ref{View selection and resource allocation}, we have the following relationship between Problem~\ref{View selection and resource allocation} and Problem~\ref{view selection and time allocation}.
\begin{lemma}[Relationship between Problem~\ref{View selection and resource allocation} and Problem~\ref{view selection and time allocation}] \label{relationship between 1 and 4} $E^*=E^\star$, $\mathbf{t}^*=\mathbf{t}^\star$, $\mathbf{y}^*=\mathbf{y}^\star$, where $E^*$ and $(\mathbf{t}^*,\mathbf{y}^*)$ are the optimal value and optimal solution of Problem~\ref{view selection and time allocation}, and $E^\star$ and $(\mathbf{t}^\star,\mathbf{y}^\star)$ are the optimal value and optimal solution of Problem~\ref{View selection and resource allocation}.
	\begin{proof}
		Please refer to Appendix C.
	\end{proof}
\end{lemma}

Recall that $\mathbf{x}^\star$ can be determined by $\mathbf{y}^\star$ according to Lemma~\ref{optimality properties}. Thus, by Lemma~\ref{relationship between 1 and 4}, we can obtain an optimal solution of Problem~\ref{View selection and resource allocation} by solving Problem~\ref{view selection and time allocation}. Note that Problem~\ref{view selection and time allocation} is a mixed discrete-continuous optimization problem and is in general NP-hard. In the following, we obtain low-complexity suboptimal solutions of Problem~\ref{view selection and time allocation}, based on which, we can obtain suboptimal solutions of Problem~\ref{View selection and resource allocation}.

\subsection{Suboptimal Solution based on Continuous Relaxation}
By relaxing the discrete constraint in (\ref{binary constraint y}) to:
\begin{equation}
y_{k,v} \in [0,1], \quad k\in \mathcal{K},\ v \in \overline{\mathcal{V}}, \label{relaxed y}
\end{equation}
we can obtain the following continuous relaxation of Problem~\ref{view selection and time allocation}.
\begin{problem}[Continuous Relaxation of Problem~\ref{view selection and time allocation}] \label{relaxed problem}
	\begin{align}
	&\min_{\mathbf{t},\mathbf{y}}~ E_1(\mathbf{t},\mathbf{y})+E_2(\mathbf{y})+E_3(\mathbf{y}) \notag\\
	&\text{ s.t.} \quad (\ref{Right constraint}),(\ref{Left constraint}),(\ref{t>=0}),(\ref{time constraint}),(\ref{relaxed y}). \notag
	\end{align}
\end{problem}
Let $(\mathbf{t}^\dagger,\mathbf{y}^\dagger)$ denote the optimal solution of Problem~\ref{relaxed problem}.

It is clear that Problem~\ref{relaxed problem} is convex and can be solved efficiently using standard convex optimization techniques. However, the optimal solution $\mathbf{y}^\dagger$ of Problem~\ref{relaxed problem} is usually not binary and hence not in the feasible set of Problem~\ref{view selection and time allocation}. Based on $\mathbf{y}^\dagger$, we can construct a feasible solution $\overline{\mathbf{y}}^\dagger \triangleq (\overline{y}_{k,v}^\dagger)_{k \in \mathcal{K}, v\in \overline{\mathcal{V}}}$ of Problem~\ref{view selection and time allocation}, as shown in Step~2--Step~8 of Algorithm~2. Based on $\overline{\mathbf{y}}^\dagger$, we can obtain $\overline{\mathbf{x}}^\dagger \triangleq (\overline{x}_v^\dagger)_{v \in \overline{\mathcal{V}}}$ according to Lemma~\ref{optimality properties} (i), as shown in Step~9 of Algorithm~2. Based on $\overline{\mathbf{x}}^\dagger$ and $\overline{\mathbf{y}}^\dagger$, we then compute $\overline{\mathbf{p}}^\dagger \triangleq (\overline{p}_v)_{v \in \overline{\mathcal{V}}}$ and $\overline{\mathbf{t}}^\dagger \triangleq (\overline{t}_v^\dagger)_{v\in \overline{\mathcal{V}}}$ according to Lemma~\ref{subproblem}, as shown in Step~10 of Algorithm~2. $(\overline{\mathbf{x}}^\dagger,\overline{\mathbf{y}}^\dagger,\overline{\mathbf{p}}^\dagger,\overline{\mathbf{t}}^\dagger)$ serves as a suboptimal solution of Problem~\ref{View selection and resource allocation}. The details are summarized in Algorithm~2.
\begin{algorithm}[!t]
	\caption{Suboptimal Solution of Problem~\ref{View selection and resource allocation} based on Continuous Relaxation}
	\textbf{Output} $(\overline{\mathbf{x}}^\dagger,\overline{\mathbf{y}}^\dagger,\overline{\mathbf{p}}^\dagger,\overline{\mathbf{t}}^\dagger)$.
	\begin{algorithmic}[1]\label{relaxed algorithm}
	\STATE Obtain $(\mathbf{t}^\dagger,\mathbf{y}^\dagger)$ by solving Problem~\ref{relaxed problem}.
	\FOR{$k\in \mathcal{K}$}
	\IF{$y^\dagger_{k,r_k}>y_{k,v}^\dagger$ for all $v\in \overline{\mathcal{V}}$ and $v \neq r_k$}
	\STATE Set $\overline{y}^\dagger_{k,r_k}=1$ and $\overline{y}^\dagger_{k,v}=0$ for all $v \in \overline{\mathcal{V}}$ and $v\neq r_k$.
	\ELSE
	\STATE Set $\overline{y}^\dagger_{k,v^+_{\max}}=\overline{y}^\dagger_{k,v^-_{\max}}=1$ and $\overline{y}^\dagger_{k,v}=0$ for all $v\in \overline{\mathcal{V}}$ and $v\neq v^+_{\max}, v^-_{\max}$, where $v^+_{\max}=\argmax_{v:v\in \overline{\mathcal{V}}_{r_k}^+} y^\dagger_{k,v}$ and $v^-_{\max}=\argmax_{v:v\in \overline{\mathcal{V}}_{r_k}^-} y^\dagger_{k,v}$.
	\ENDIF
	\ENDFOR
	\STATE Set $\overline{x}^\dagger_v = \max_{k \in \mathcal{K}} \overline{y}^\dagger_{k,v}$ for all $v \in \overline{\mathcal{V}}$.
	\STATE For given $\overline{\mathbf{x}}^\dagger$ and $\overline{\mathbf{y}}^\dagger$, compute $(\overline{\mathbf{p}}^\dagger,\overline{\mathbf{t}}^\dagger)$ based on Lemma~\ref{subproblem}.
	\end{algorithmic}
\end{algorithm}
\subsection{Suboptimal Solution based on DC Programming}
The discrete constraint in (\ref{binary constraint y}) can be equivalently transformed to (\ref{relaxed y}) and
\begin{equation}
y_{k,v}(1-y_{k,v})\leq 0, \quad k\in\mathcal{K},~v\in \overline{\mathcal{V}}.\label{transformed binary <y}
\end{equation}
Then, Problem~\ref{view selection and time allocation} can be equivalently transformed to the following problem.
\begin{problem}[DC Problem of Problem~\ref{view selection and time allocation}]\label{DC transform}
	\begin{align}
	&\min_{\mathbf{t},\mathbf{y}}~ E_1(\mathbf{t},\mathbf{y})+E_2(\mathbf{y})+E_3(\mathbf{y}) \notag\\
	&\text{ s.t.} \quad (\ref{Right constraint}),(\ref{Left constraint}),(\ref{t>=0}),(\ref{time constraint}),(\ref{relaxed y}),(\ref{transformed binary <y}). \notag
	\end{align}
\end{problem}

Note that the constraint function in (\ref{transformed binary <y}) is concave. Thus, Problem~\ref{DC transform} is a difference of convex (DC) problem \cite{lipp2016variations}. In the following, we adopt the DC method in \cite{le2012exact} to obtain a stationary point of Problem~\ref{DC transform}. First, we approximate Problem~\ref{DC transform} by disregarding the constraint in (\ref{transformed binary <y}) and adding to the objective function a penalty for violating the constraint in (\ref{transformed binary <y}).

\begin{problem}[Penalized Problem of Problem~\ref{DC transform}]\label{Penalty relaxed problem}
\begin{align}
&\min_{\mathbf{t},\mathbf{y}}~ E_1(\mathbf{t},\mathbf{y})+E_2(\mathbf{y})+E_3(\mathbf{y})+\rho P(\mathbf{y}) \notag\\
&~\text{s.t.}\quad(\ref{Right constraint}),(\ref{Left constraint}),(\ref{t>=0}),(\ref{time constraint}),(\ref{relaxed y}), \notag
\end{align}
\end{problem}
where the penalty parameter $\rho>0$ and the penalty function $P(\mathbf{y})$ is given by
$P(\mathbf{y})=\sum_{k\in \mathcal{K}}\sum_{v\in \overline{\mathcal{V}}} y_{k,v}(1-y_{k,v})$.

Note that the objective function of Problem~\ref{DC transform} is Lipschitz continuous and the feasible set of Problem~\ref{DC transform} is a nonempty bounded polyhedral convex set. Thus, there exists $\rho_0>0$ such that for all $\rho>\rho_0$, Problem~\ref{Penalty relaxed problem} is equivalent to Problem~\ref{DC transform} \cite{le2012exact}. Now, we solve Problem~7 instead of Problem~6 by using the DC algorithm in \cite{lipp2016variations}. The main idea is to iteratively solve a sequence of convex approximations of Problem~\ref{Penalty relaxed problem}, each of which is obtained by linearizing the penalty function $P(\mathbf{y})$ in the objective function of Problem~\ref{Penalty relaxed problem}. Specifically, we have:
\begin{problem}\textit{(Convex Approximation of Problem~\ref{Penalty relaxed problem} at $i$-th Iteration):}\label{Convexified Penalty Problem}
\begin{align}
(\mathbf{t}^{(i)},\mathbf{y}^{(i)}) \triangleq
\arg&\min_{\mathbf{t},\mathbf{y}} E_1(\mathbf{t},\mathbf{y})+E_2(\mathbf{y})+E_3(\mathbf{y})+\rho\hat{P}\left(\mathbf{y};\mathbf{y}^{(i-1)}\right) \notag\\
&~\text{s.t.}\quad (\ref{Right constraint}),(\ref{Left constraint}),(\ref{t>=0}),(\ref{time constraint}),(\ref{relaxed y}), \notag
\end{align}
\end{problem}
where
\begin{align}
\hat{P}\left(\mathbf{y};\mathbf{y}^{(i-1)}\right)&\triangleq P\left(\mathbf{y}^{(i-1)}\right)+\nabla P\left(\mathbf{y}^{(i-1)}\right)^T \left(\mathbf{y}-\mathbf{y}^{(i-1)}\right) \notag \\
&=\sum_{k \in \mathcal{K}} \sum_{v\in \overline{\mathcal{V}}} \left( \left(1-2y_{k,v}^{(i-1)}\right)y_{k,v}+\left(y_{k,v}^{(i-1)}\right)^2 \right). \notag
\end{align}
Here, $\mathbf{y}^{(i-1)}$ denotes the solution of Problem~\ref{Convexified Penalty Problem} at the $(i-1)$-th iteration.

It has been shown that the DC algorithm can obtain a stationary point of Problem~\ref{Penalty relaxed problem} \cite{lipp2016variations}, denoted by $(\mathbf{t}^\dagger,\mathbf{y}^\dagger)$, with slight abuse of notation. Due to the equivalence among Problems~\ref{view selection and time allocation}, \ref{DC transform} and \ref{Penalty relaxed problem}, we know that $(\mathbf{t}^\dagger,\mathbf{y}^\dagger)$ is a feasible solution of Problem~\ref{view selection and time allocation}. Similarly, based on $\mathbf{y}^\dagger$, we can obtain $\mathbf{x}^\dagger \triangleq (x_v^\dagger)_{v \in \overline{\mathcal{V}}}$ according to Lemma~\ref{optimality properties} (i), as shown in Step~7 of Algorithm~3. Based on $\mathbf{x}^\dagger$ and $\mathbf{y}^\dagger$, we then compute $\mathbf{p}^\dagger \triangleq (p_v^\dagger)_{v \in \overline{\mathcal{V}}}$ according to Lemma~\ref{subproblem}, as shown in Step~8 of Algorithm~3. $(\mathbf{x}^\dagger,\mathbf{y}^\dagger,\mathbf{p}^\dagger,\mathbf{t}^\dagger)$ serves as a suboptimal solution of Problem~\ref{View selection and resource allocation}. The details are summarized in Algorithm~3.
\begin{algorithm}[!t]
	\caption{Suboptimal Solution of Problem~\ref{View selection and resource allocation} based on DC Programming}
	\textbf{Output} $(\mathbf{x}^\dagger,\mathbf{y}^\dagger,\mathbf{p}^\dagger,\mathbf{t}^\dagger)$
	\begin{algorithmic}[1] \label{DC Algorithm}
		\STATE Find an initial feasible point $(\mathbf{t}^{(0)},\mathbf{y}^{(0)})$, choose a sufficiently large $\rho$, and set $i=0$.
		\REPEAT
		\STATE Set $i=i+1$.
		\STATE Obtain $(\mathbf{t}^{(i)},\mathbf{y}^{(i)})$ of Problem~\ref{Convexified Penalty Problem} using standard convex optimization techniques.
		\UNTIL convergence criteria is met.
		\STATE Set $\mathbf{y}^\dagger = \mathbf{y}^{(i)}$ and $\mathbf{t}^\dagger = \mathbf{t}^{(i)}$.
		\STATE Set $x_v^\dagger =\max_{k\in\mathcal{K}} y_{k,v}^\dagger$ for all $v\in \overline{\mathcal{V}}$.
		\STATE For given $\mathbf{x}^\dagger$ and $\mathbf{y}^\dagger$, compute $\mathbf{p}^\dagger$ based on Lemma~\ref{subproblem}.
		
	\end{algorithmic}
\end{algorithm}

\section{Simulation}
In the simulation, we set $\beta=3$, $R=10$Mbit/s, $E_b=5\times10^{-7}$Joule, $E_{\text{u},k}=5\times10^{-7}$Joule for all $k \in \mathcal{K}$, $\mathcal{V}=\{1,2,3,4,5\}$, $\overline{\mathcal{V}} =\{1,1.1,1.2,\cdots,5\}$ (i.e., $Q=10$), $\Delta=1$ and $n_0=Bk_BT_0$, where $k_B=1.38\times10^{-23}$Joule/Kelvin is the Boltzmann constant and $T_0=300$Kelvin is the temperature. For all $k \in \mathcal{K}$, we assume channel power $h_k$ follows Rayleigh fading with mean $10^{-3}$ (which is to reflect path loss). In addition, for all $k\in\mathcal{K}$, we assume view request $r_k$ follows the uniform distribution over $\overline{\mathcal{V}}$. We generate 100 random independent channel powers and view requests for all users, and evaluate the average performance over these realizations. We use Matlab software and CVX toolbox to implement Algorithms~1, 2 and 3.
\subsection{Comparison between Optimal and Suboptimal Solutions}
\begin{figure}[!ht]
	\subfigure[Weighted sum energy consumption vs $K$.]{ 
	\begin{minipage}{0.47\linewidth}
		\centering
		\includegraphics[width=\linewidth,height=6cm]{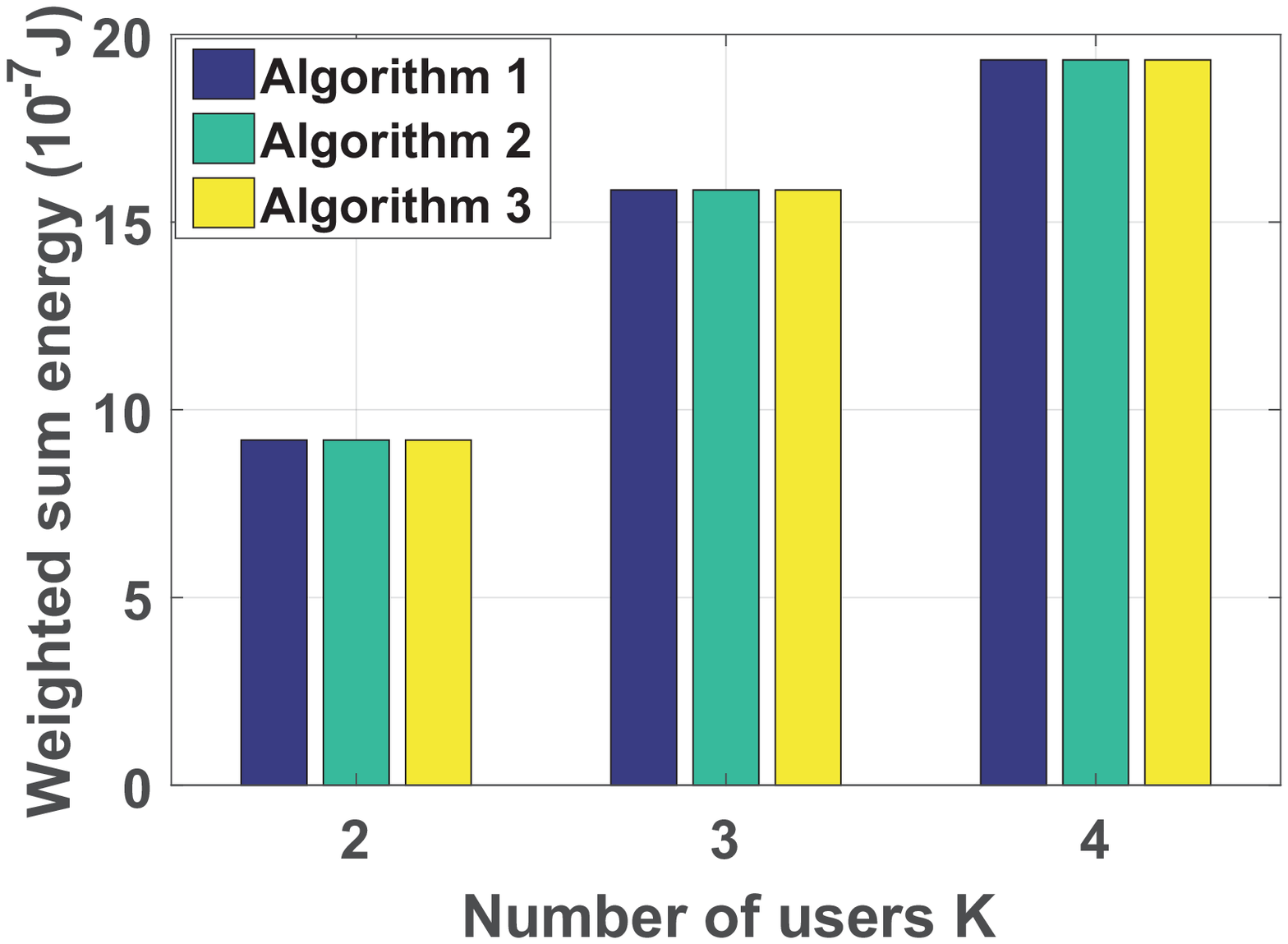}
	\end{minipage}
}
\subfigure[Computation time vs $K$.]{
	\begin{minipage}{0.47\linewidth}
		\centering
		\includegraphics[width=\linewidth,height=6cm]{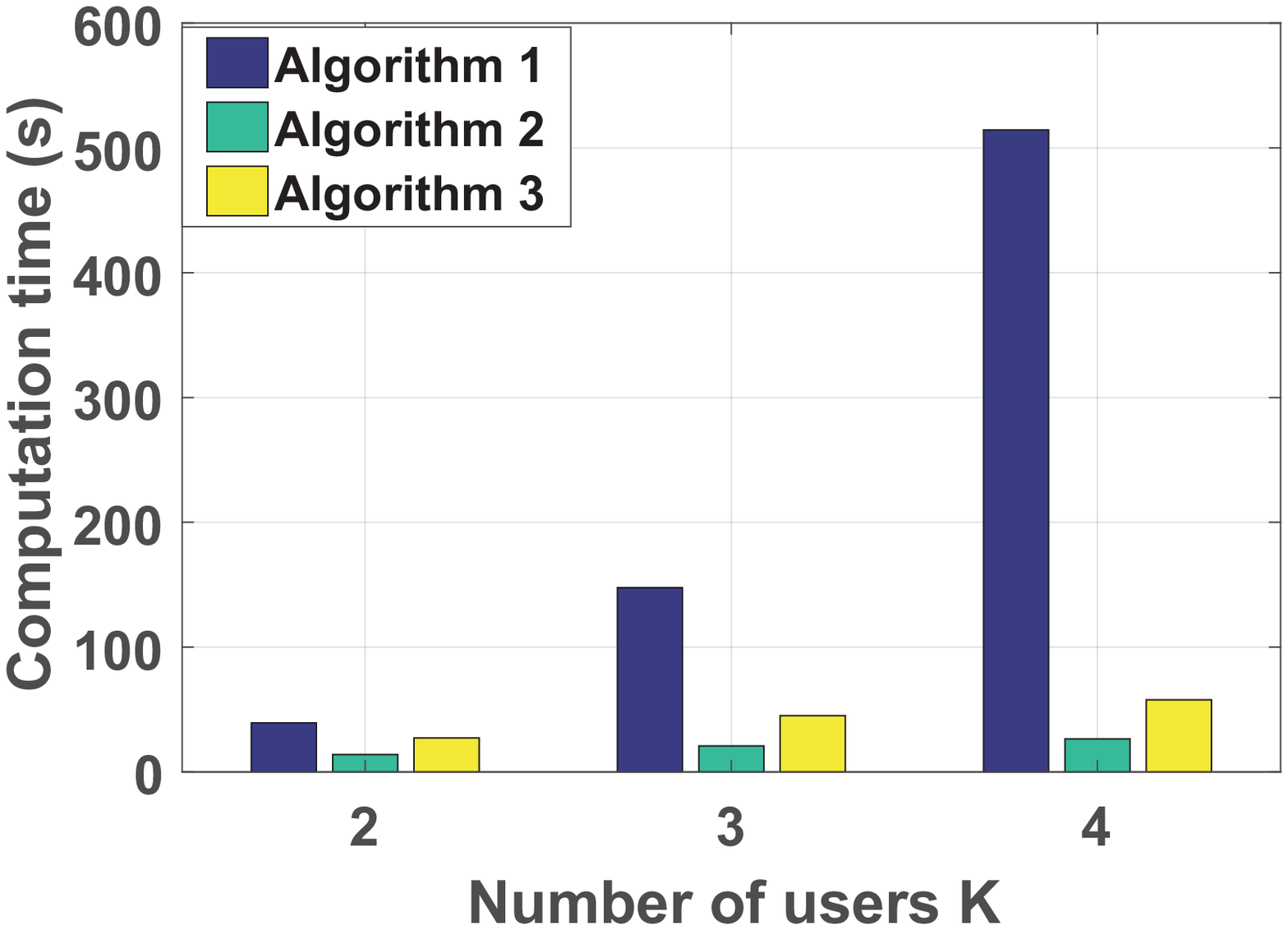}
	\end{minipage}
}
\caption{Comparison between the optimal and suboptimal solutions at $B=10$MHz and $T=100$ms.}
\label{simulation_K}
\end{figure}

In this part, we use a numerical example for a small\footnote{Note that the computational complexity of Algorithm~1 is not acceptable when $K$ is larger.} $K$ to compare the optimal solution (obtained using Algorithm~1) and the proposed suboptimal solutions (obtained using Algorithm~2 and Algorithm~3). Fig.~\ref{simulation_K} illustrates the weighted sum energy consumption and computation time versus the number of users, respectively. From Fig. \ref{simulation_K} (a), we can see that the weighted sum energy of each suboptimal solution is very close to that of the optimal solution. From Fig. \ref{simulation_K} (b), we can see that the computation time of each suboptimal solution grows at a much smaller rate than the optimal solution with respect to the number of users. In addition, the computation time of the suboptimal solution based on continuous relaxation (i.e., Algorithm~2) is smaller than that of the suboptimal solution based on DC programming (i.e., Algorithm~3). This numerical example demonstrates the applicability and efficiency of the suboptimal solutions.
\subsection{Comparison between Suboptimal Solutions and Baseline Schemes}
In this part, we compare two proposed suboptimal solutions with two baseline schemes. Baseline~1 considers view synthesis at the server but does not consider view synthesis at the user side. More specifically, for all $k\in\mathcal{K}$, $x_v=1$ if $v=r_k$, and $x_v=0$ otherwise; for all $k \in \mathcal{K}$, $y_{k,v}=1$ if $v=r_k$ and $y_{k,v}=0$ otherwise. Baseline~2 considers view synthesis at the user side but does not consider view synthesis at the server. More specifically, for all $k \in \mathcal{K}$, $x_v=1$ if $v= \lfloor r_k \rfloor$ or $\lceil r_k \rceil$, and $x_v=0$ otherwise; for all $k \in \mathcal{K}$, $y_{k,v}=1$  if $v= \lfloor r_k \rfloor$ or $\lceil r_k \rceil$ and $y_{k,v}=0$ otherwise. Based on $\mathbf{x}$ and $\mathbf{y}$, both baseline schemes adopt the optimal power and time allocation according to Lemma~\ref{subproblem}, as in the proposed solutions.

\begin{figure}[t]
	\subfigure[Weighted sum energy consumption vs $B$ at $T=100$ms.]
	{ 
	\begin{minipage}{0.47\linewidth}
			\centering
			\includegraphics[width=\linewidth,height=6cm]{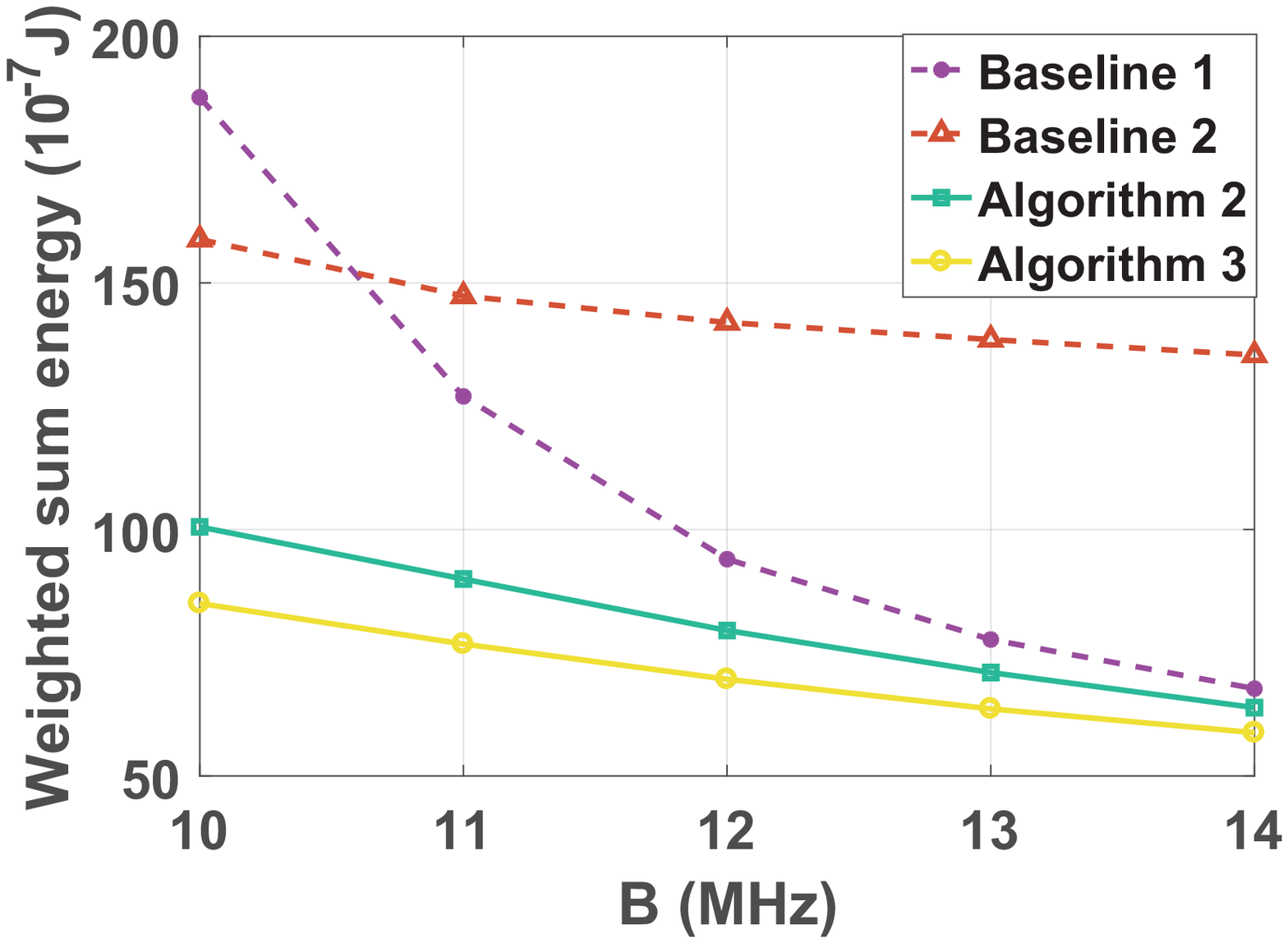}
	
		\end{minipage}
	}
\hspace{-2mm}
	\subfigure[Weighted sum energy consumption vs $T$ at $B=10$MHz.]
	{ 
		\begin{minipage}{0.47\linewidth}
			\centering
			\includegraphics[width=\linewidth,height=6cm]{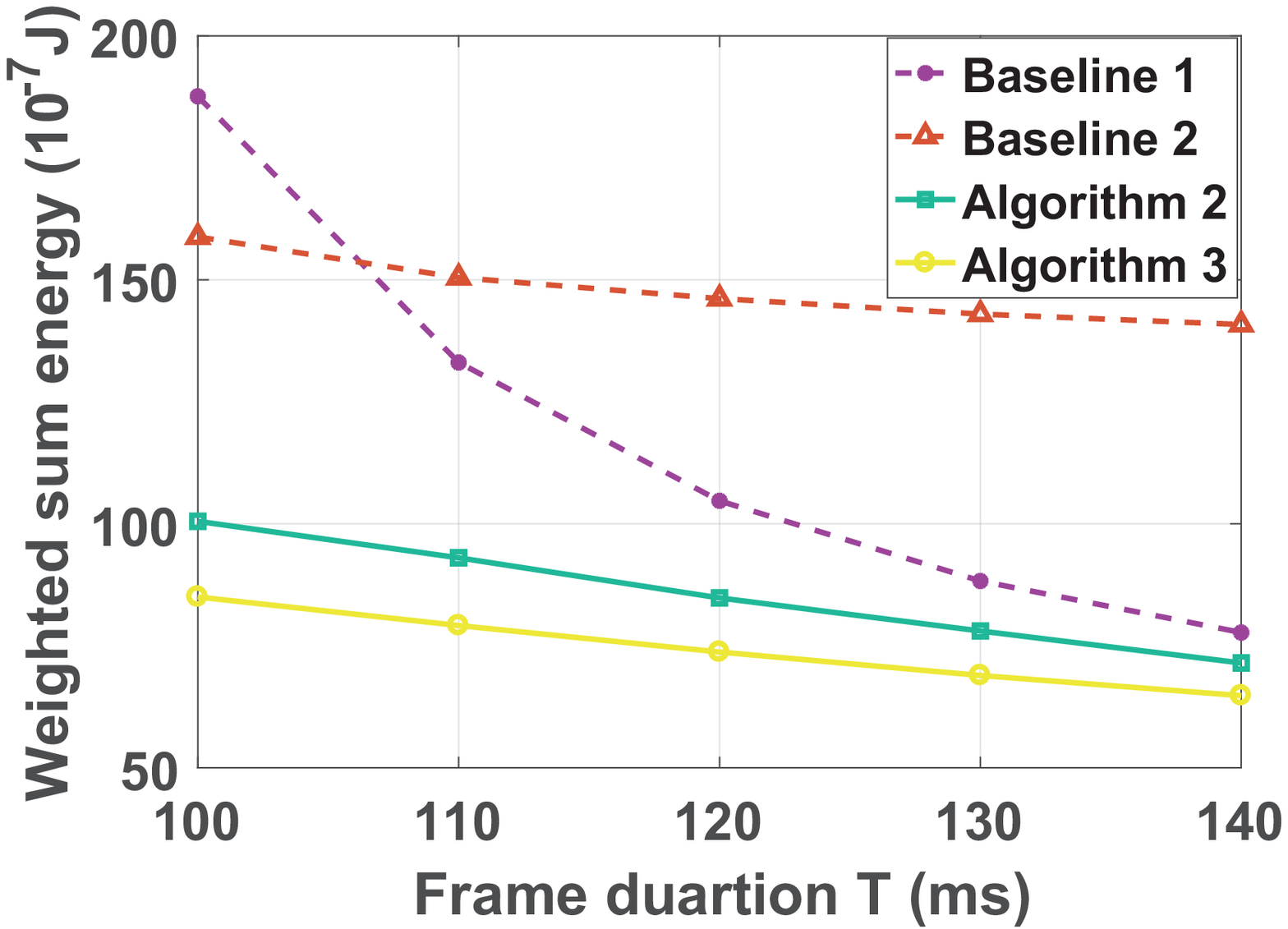}
			
		\end{minipage}
	}
\vspace{2mm}
\caption{Comparison between the suboptimal solutions and two baseline schemes at $K=10$.}
\label{simulation_B}
\end{figure}

Fig.~\ref{simulation_B} illustrates the weighted sum energy consumption versus the bandwidth and frame duration, respectively. From Fig.~\ref{simulation_B}, we can see that the weighted sum energy consumption of each scheme decreases as the bandwidth or frame duration increases. In addition, we can see that when the bandwidth or frame duration is small, Baseline~2 outperforms Baseline~1, and when the bandwidth or frame duration is large, Baseline~1 outperforms Baseline~2. The reasons are as follows. (i) Baseline~1 (view synthesis at the server) incurs smaller weighted synthesis energy consumption than Baseline~2 (view synthesis at the user side), as $\beta E_{\text{u},k}>E_{b}$ for all $k \in \mathcal{K}$. (ii) Baseline~1 has higher transmission energy consumption than Baseline~2, as Baseline~1 has fewer multicast opportunities and transmits more views. (iii) As the bandwidth or frame duration increases, the transmission energy consumption decreases but the synthesis energy consumption does not change. Finally, from Fig.~\ref{simulation_B}, we see that the two suboptimal solutions outperform the two baseline schemes, demonstrating the advantage of the proposed solutions in making full use of view synthesis-enabled multicast opportunities.

\section{Conclusion}
In this paper, we considered energy-efficient MVV transmission from a server to multiple users in a wireless network using TDMA. View synthesis was allowed at the server and each user to create  multicast opportunities for further improving transmission efficiency and reducing energy consumption. Specifically, we first established a mathematical model to characterize the impact of view synthesis on multicast opportunities and energy consumption. Then, we considered the optimization of view selection, transmission time and power allocation to minimize the weighted sum energy consumption for view transmission and synthesis, which is a challenging mixed discrete-continuous optimization problem. We proposed an optimal algorithm  with reduced computational complexity and two low-complexity suboptimal algorithms to solve the problem. Finally, numerical results demonstrate the advantage of the proposed solutions. To the best of our knowledge, this is the first work providing optimization-based solutions for energy-efficient MVV transmission by exploiting view synthesis-enabled multicast opportunities in multiuser wireless networks.


\section*{Appendix A: Proof of Lemma~1}
First, we transform Problem~\ref{sub-problem for resource allocation} into an equivalent convex problem. For all $v\in\overline{\mathcal{V}}$, consider two cases. Case (i): $\sum_{k\in\mathcal{K}} y_{k,v}=0$. In this case, by contradiction, we can easily show that setting $p_v=0$ and $t_v=0$ will not loss optimality. Case (ii): $\sum_{k\in\mathcal{K}} y_{k,v}>0$. In this case, by (8), we know that $p_v>0$ and \begin{equation}\label{t>0}
t_v>0, \quad v \in \overline{\mathcal{V}}.
\end{equation}
Thus, (8) can be rewritten as
\begin{equation}\label{inequality between t and p}
p_v \geq \frac{n_0}{h_k} \left( 2^{\frac{y_{k,v}RT}{t_vB}}-1 \right),\quad k\in\mathcal{K},\ v \in \overline{\mathcal{V}}.
\end{equation}
It is clear that setting
\begin{equation}\label{relationship}
p_v =\frac{n_0}{h_{v,\min}(\mathbf{y}_v)} \left( 2^{\frac{y_{k,v}RT}{t_vB}}-1 \right) \quad k\in\mathcal{K},\ v \in \overline{\mathcal{V}}
\end{equation}
will not lose optimality. Therefore, we can equivalently transform Problem~\ref{sub-problem for resource allocation} into the following problem.

\begin{problem}[Equivalent Problem of Problem~\ref{sub-problem for resource allocation}]\label{proof of transformed convex problem} For given $\mathbf{x}$ and $\mathbf{y}$ that satisfy (\ref{binary constraint x})-(\ref{x>y}), we have:
\begin{align}
E_\text{t}^\star(\mathbf{x},\mathbf{y}) \triangleq&\min_{\mathbf{t}} \quad \sum_{v \in \overline{\mathcal{V}}} \frac{n_0t_v}{h_{v,\min}(\mathbf{y}_v)} \left( 2^{\frac{y_{k,v}RT}{t_vB}}-1 \right) \notag\\
&~\text{s.t.} \quad (\ref{time constraint}),(\ref{t>0}). \notag
\end{align}
\end{problem}
It is obvious that Problem~\ref{proof of transformed convex problem} is convex and slater's condition holds. Thus strong duality holds. The Lagrange function of Problem~\ref{sub-problem for resource allocation} is given by
\begin{align}
L(\mathbf{t},\mu,\boldsymbol{\lambda})= \sum_{v \in \overline{\mathcal{V}}} \left( \frac{n_0t_v}{h_{v,\min}(\mathbf{y}_v)} \left( 2^{\frac{y_{k,v}RT}{t_vB}}-1 \right) \right) + \lambda \left( \sum_{v\in \overline{\mathcal{V}}} t_v-T \right)+\sum_{v \in \overline{\mathcal{V}}} \mu_v t_v
\end{align}
where $\boldsymbol{\mu}\triangleq (\mu_v)_{v\in \overline{\mathcal{V}}}$ and $\lambda$ are the Lagrange multipliers associated with inequality constraints $\sum_{v\in\overline{\mathcal{V}}} t_v\leq T$ and $t_v>0,\ v\in \overline{\mathcal{V}}$, respectively. Thus, we have:
\begin{align}
&\frac{\partial L(\mathbf{t},\mu,\boldsymbol{\lambda})}{\partial t_v}=\frac{n_0}{h_{v,\min}(\mathbf{y}_v)}\left(\left( 1- \frac{y_{k,v}RT\log 2 }{Bt_v}\right)2^{\frac{y_{k,v}RT}{t_vB}}-1 \right)+\mu_v+\lambda,\quad v\in \overline{\mathcal{V}}.
\end{align}
Since strong duality holds, primal optimal $\mathbf{t}^\star$ and dual optimal $\mu^\star$ and $\boldsymbol{\lambda}^\star$ satisfy KKT conditions: (i) primal constraints (7) and (18); (ii) dual constraints $\mu\geq0$ and $\lambda_v\geq0$ for all $v\in\overline{\mathcal{V}}$; (iii) complementary slackness $\mu \left( \sum_{v\in \overline{\mathcal{V}}} t_v-T \right)=0$ and $\lambda_v t_v=0$ for all $v\in\overline{\mathcal{V}}$; and (iv) $\frac{\partial L(\mathbf{t},\mu,\boldsymbol{\lambda})}{\partial t_v}=0$. By (i)-(iv), we can get the closed-form solution:
\begin{align}
&t_v^\star(\mathbf{x},\mathbf{y})= \begin{cases}
\frac{RT \log2}{B\left( W(\frac{\lambda^\star(\mathbf{x},\mathbf{y}) h_{v,\min}(\mathbf{y}_v)}{n_0 e}-\frac{1}{e})+1\right)}, & \sum_{k \in \mathcal{K}} y_{k,v}>0\\
0, & \sum_{k \in \mathcal{K}} y_{k,v}=0 \end{cases}
,\ v \in \overline{\mathcal{V}}, \notag \\
&p_v^\star(\mathbf{x},\mathbf{y})=\begin{cases}
\frac{n_0}{h_{v,\min}(\mathbf{y}_v)}\left( 2^{\frac{RT}{Bt_v^\star(\mathbf{x},\mathbf{y})}}-1\right), & \sum_{k \in \mathcal{K}} y_{k,v}>0  \notag\\
0, & \sum_{k \in \mathcal{K}} y_{k,v}=0
\end{cases}
,\ v \in \overline{\mathcal{V}},
\end{align}
where $\mathbf{y}_v \triangleq (y_{k,v})_{k\in \mathcal{K}}$, and $\lambda^\star(\mathbf{x},\mathbf{y})$ satisfies
\begin{equation}
\sum_{v\in \overline{\mathcal{V}}} \frac{RT\log 2}{B\left( W(\frac{\lambda^\star(\mathbf{x},\mathbf{y}) h_{v,\min}(\mathbf{y}_v)}{n_0 e}-\frac{1}{e})+1\right)}x_v=T. \notag
\end{equation}
\section*{Appendix B: Proof of Lemma~2}
\subsection{Proof of Statement (i)}
We prove Statement~(i) by contradiction. Suppose there exists $v\in\overline{\mathcal{V}}$ such that $x_v^\star\neq \max_{k\in \mathcal{K}} y_{k,v}^\star$. By (5), this implies $x_v^\star> \max_{k\in \mathcal{K}} y_{k,v}^\star$. By (1) and (2), we know $x_v^\star=1$ and $y_{k,v}^\star=0$ for all $k\in \mathcal{K}$. Construct $x_v^\dagger=0$. Note that $x_v^\dagger=0$ and $y_{k,v}^\star=0$ for all $k\in\mathcal{K}$ satisfy (1) and (5). In addition, the objective function of Problem~\ref{View selection and resource allocation} $E(\mathbf{t},\mathbf{p},\mathbf{x})$ increases with $x_v$, and the constraints in (3),(6),(7),(8) are independent of $x_v$. Therefore, $x_v^\dagger=0$ and $y_{k,v}^\star=0$ for all $k\in\mathcal{K}$ lead to a smaller objective value. This contradicts the assumption. Therefore, by contradiction, we can prove Statement~(i).
\subsection{Proof of Statement (ii)}

We prove Statement(ii) by contradiction. Suppose that there exist $k_0 \in \mathcal{K}$ and $v_0 \notin U_{k_0}$ such that $ y^\star_{k_0,v_0} = 1$. First it is easy to show that if $v_0 \notin \mathcal{V}_{k_0}^- \cup \mathcal{V}_{k_0}^+$, it is clear that $y_{k,v_0}=0$ leads to a smaller objective value. In the following, we consider $v_0 \in \mathcal{V}_{k_0}^+$. The argument for $v_0 \in \mathcal{V}_{k_0}^-$ is similar. Define $\mathcal{K}_{v_0} \triangleq \{k\in \mathcal{K}~|y_{k,v_0} = 1\}$ , $v_1 \triangleq \max_{k \in \mathcal{K}_{v_0}} f_{v_0}(r_k)$, where
\begin{equation*}
f_{v_0}(r_k)\triangleq \begin{cases}
r_k & r_k<v_0\\
r_k-\Delta & r_k>v_0
\end{cases},
\end{equation*}
and $\mathcal{M}_{v_0}\triangleq\{k\in \mathcal{K}_{v_0} | r_k=v_1\}$. By (\ref{Right constraint}) and $y_{k,v_0}^\star=1 $, we know $y_{k,v_1}^\star=0$ for all $k \in \mathcal{K}_{v_0}$. Based on Lemma~1 and Lemma~2 (i), we can obtain $\mathbf{x}^{\star}$, $\mathbf{t}^{\star}$ and $\mathbf{p}^{\star}$. Let $E^{\star}=E(\mathbf{t}^{\star},\mathbf{p}^{\star},\mathbf{x}^{\star})$ and $\lambda^{\star}\triangleq \lambda^\star(\mathbf{x}^\star,\mathbf{y}^\star)$. Next, we construct the solution $\mathbf{y}^{\dagger}\triangleq(y_{k,v}^\dagger)_{k\in\mathcal{K},v\in\overline{\mathcal{V}}}$ where $y_{k,v_1}=1$ and $y_{k,v_0}=0$ for all $k \in \mathcal{K}_{v_0}$.  Similarly, we can obtain $\mathbf{x}^{\dagger}$, $\mathbf{t}^{\dagger}$ and $\mathbf{p}^{\dagger}$. Let $E^{\dagger}=E(\mathbf{t}^{\dagger},\mathbf{p}^{\dagger},\mathbf{x}^{\dagger})$ and $\lambda^{\dagger}\triangleq\lambda^\star(\mathbf{x}^\dagger,\mathbf{y}^\dagger)$.
In the following, we prove $E^{\star}\geq E^{\dagger}$ by considering two cases.
\begin{case}
    $v_1\in \{r_k~ |~ k\in \mathcal{K}\}$. In this case, we have
	\begin{align}
	E^{\star}-E^{\dagger}&=E_\text{t}(\mathbf{t}^{\star},\mathbf{p}^{\star})+E_{\text{s}}^{(b)}(\mathbf{x}^{\star})+\beta E_{\text{s}}^{(u)}(\mathbf{x}^{\star}) - E_\text{t}(\mathbf{t}^{\dagger},\mathbf{p}^{\dagger})-E_{\text{s}}^{(b)}(\mathbf{x}^{\dagger})-\beta E_{\text{s}}^{(u)}(\mathbf{x}^{\dagger}), \notag \\
	&=\left(E_\text{t}(\mathbf{t}^{\star},\mathbf{p}^{\star})-E_\text{t}(\mathbf{t}^{\dagger},\mathbf{p}^{\dagger})\right)+\left(E_{\text{s}}^{(b)}(\mathbf{x}^{\star})-E_{\text{s}}^{(b)}(\mathbf{x}^{\dagger})\right)+\beta\left(E_{\text{s}}^{(u)}(\mathbf{x}^{\star})-E_{\text{s}}^{(u)}(\mathbf{x}^{\dagger})\right) \notag \\
	&=\left(E_\text{t}(\mathbf{t}^{\star},\mathbf{p}^{\star})-E_\text{t}(\mathbf{t}^{\dagger},\mathbf{p}^{\dagger})\right)+\sum_{k \in \mathcal{K}} \left(y^{\dagger}_{k,r_k}- y^{\star}_{k,r_k}\right)E_{u,k}+(x_{v_0}^{\star}+x_{v_1}^{\star}-x_{v_0}^{\dagger}-x_{v_1}^{\dagger})E_b\notag\\
	&\overset{(a)}{\geq} E_\text{t}(\mathbf{t}^{\star},\mathbf{p}^{\star})-E_\text{t}(\mathbf{t}^{\dagger},\mathbf{p}^{\dagger}) + \beta \sum_{k\in \mathcal{M}} E_{u,k}-E_b, \notag\\
	&\overset{(b)}{\geq} E_\text{t}(\mathbf{t}^{\star},\mathbf{p}^{\star})-E_\text{t}(\mathbf{t}^{\dagger},\mathbf{p}^{\dagger}) ,
	\end{align}
	where (a) is due to $x_{v_0}^{\dagger}=0$, $x_{v_0}^{\star}+x_{v_1}^{\star}-x_{v_1}^{\dagger}\geq -1$,  $y_{k,v_1}^{\star}=y_{k,r_k}^{\star}=0$, $y_{k,v_0}^{\dagger}=y_{k,r_k}^{\dagger}=1$ for all $k\in\mathcal{M}$, $y_{k,r_k}^{\star}=y_{k,r_k}^{\dagger}$ for all $k \notin \mathcal{M}$ and $| \mathcal{M}|\geq 1$, and (b) is due to $\beta E_{u,k}-E_b\geq 0$ for all $k \in \mathcal{K}$. To show $E^{\star}\geq E^{\dagger}$, it remains to show $E_\text{t}(\mathbf{t}^{\star},\mathbf{p}^{\star})-E_\text{t}(\mathbf{t}^{\dagger},\mathbf{p}^{\dagger})\geq 0$. First, we have	
	\begin{align}
	\sum_{v\in \overline{\mathcal{V}}} t_{v}(\lambda^{\star},\mathbf{y}^{\star})-\sum_{v\in \overline{\mathcal{V}}} t_{v}(\lambda^{\dagger},\mathbf{y}^{\star})&=
	T -\sum_{v\in \overline{\mathcal{V}}} t_v(\lambda^{\dagger},\mathbf{y}^{\star})) \notag\\
	&=T- \sum_{v\in \overline{\mathcal{V}} \setminus \{v_0,v_1\}} t_v(\lambda^{\dagger},\mathbf{y}^{\star}) + t_{v_0}(\lambda^{\dagger},\mathbf{y}^{\star}) +t_{v_1}(\lambda^{\dagger},\mathbf{y}^{\star})\notag \\
	&\overset{(c)}{=} T-\left(T - t_{v_0}(\lambda^{\dagger},\mathbf{y}^{\dagger}) -t_{v_1}(\lambda^{\dagger},\mathbf{y}^{\dagger})+t_{v_0}(\lambda^{\dagger},\mathbf{y}^{\star}) +t_{v_1}(\lambda^{\dagger},\mathbf{y}^{\star})\right) \notag \\
	&\overset{}{=} t_{v_0}(\lambda^{\dagger},\mathbf{y}^{\dagger})+t_{v_1}(\lambda^{\dagger},\mathbf{y}^{\dagger})-t_{v_0}(\lambda^{\dagger},\mathbf{y}^{\star}) -t_{v_1}(\lambda^{\dagger},\mathbf{y}^{\star})\notag\\
	&\overset{(d)}{\leq} 0
	\end{align}	
	where (c) is due to $h_{v,\min}^{\star}=h_{v,\min}^{\dagger}$ for all $v\in \overline{\mathcal{V}} \setminus \{v_0,v_1\}$, and (d) is due to the fact that $h_{v_1,\min}(\mathbf{y}^{\dagger}) = \min(h_{v_1,\min}(\mathbf{y}^{\star}),h_{v_0,\min}(\mathbf{y}^{\star}))$ and $t_v(\lambda,\mathbf{y})$ is strictly decreasing function of $h_{v,\min}(\mathbf{y}_v)$. We have $t_{v_1}(\lambda^{\dagger},\mathbf{y}^{\dagger}) = \max(t_{v_1}(\lambda^{\dagger},\mathbf{y}^{\star}),t_{v_0}(\lambda^{\dagger},\mathbf{y}^{\star}))$ and $t_{v_0}(\mathbf{y}^{\dagger}) = 0$. Thus we have
	\begin{align}
	 &t_{v_0}(\lambda^{\dagger},\mathbf{y}^{\dagger})+t_{v_1}(\lambda^{\dagger},\mathbf{y}^{\dagger})-t_{v_0}(\lambda^{\dagger},\mathbf{y}^{\star})  -t_{v_1}(\lambda^{\dagger},\mathbf{y}^{\star}) \notag \\
	 &= \max(t_{v_1}(\lambda^{\dagger},\mathbf{y}^{\star}),t_{v_0}(\lambda^{\dagger},\mathbf{y}^{\star})) -t_{v_0}(\lambda^{\dagger},\mathbf{y}^{\star}) -t_{v_1}(\lambda^{\dagger},\mathbf{y}^{\star}) \leq 0
	\end{align}
	By $\sum_{v\in \overline{\mathcal{V}}} t_{v}(\lambda^{\star},\mathbf{y}^{\star})-\sum_{v\in \overline{\mathcal{V}}} t_{v}(\lambda^{\dagger},\mathbf{y}^{\star})\leq0$ and the fact that $\sum_{v\in \overline{\mathcal{V}}}t_v(\lambda,\mathbf{y}^{\star})$ is a strictly decreasing function with respect to $\lambda$, we know $\lambda^\star\geq \lambda^\dagger$. Thus, we have
	\begin{align}
	E_\text{t}(\mathbf{t}^{\star},\mathbf{p}^{\star})-E_\text{t}(\mathbf{t}^{\dagger},\mathbf{p}^{\dagger})&=\sum_{v\in \overline{\mathcal{V}}} p_v^{\star}t_{v}^{\star}-\sum_{v\in \overline{\mathcal{V}}} p_v^{\dagger}t_{v}^{\dagger} \notag \\
	&= \sum_{v\in \overline{\mathcal{V}} \setminus \{v_0,v_1\}} \left(p_v^{\star}t_{v}^{\star}-p_v^{\dagger}t_{v}^{\dagger} \right) + p_{v_0}^{\star}t_{v_0}^{\star} +p_{v_1}^{\star}t_{v_1}^{\star}-p_{v_0}^{\dagger}t_{v_0}^{\dagger}-p_{v_1}^{\dagger}t_{v_1}^{\dagger} \notag\\
	&\overset{(e)}{\geq}p_{v_0}^{\star}t_{v_0}^{\dagger} +p_{v_1}^{\star}t_{v_1}^{\star}-p_{v_0}^{\dagger}t_{v_0}^{\dagger}-p_{v_1}^{\dagger}t_{v_1}^{\dagger} \notag\\
	&\overset{(f)}{\geq}0
 	\end{align}
	where (e) is due to that for $\lambda^{\star} \geq \lambda^{\dagger}$, we have $t_{v}(\lambda^{\star},\mathbf{y}^{\star}) \leq t_{v}(\lambda^{\dagger},\mathbf{y}^{\dagger})$ for all $ v\in \overline{\mathcal{V}} \setminus \{v_0,v_1\}$. Because $p_vt_v$ is a strictly monotonicity decrease function by $t_v$, we have $E_v(\mathbf{y}^{\star}) \geq E_v(\mathbf{y}^{\dagger}), v\in \overline{\mathcal{V}} \setminus \{v_0,v_1\}$.
	
	(f) is due to that because $h_{v_1,\min}(\mathbf{y}^{\dagger}) = \min(h_{v_1,\min}(\mathbf{y}^{\star},h_{v_0,\min}(\mathbf{y}^{\star}))$, we have $p_{v_1}^{\dagger}t_{v_1}^{\dagger} \leq \max(p_{v_0}^{\star}t_{v_0}^{\star},p_{v_1}^{\star}t_{v_1}^{\star})$ and $p_{v_0}^{\dagger}t_{v_0}^{\dagger}=0$.
\end{case}

\begin{case}
		Consider $v_1\notin \{r_k~ |~ k\in \mathcal{K}\}$. First, it's easy to show that $v_0 \notin \mathcal{V}$, that's because all origin views in $[v_1,r_{k_0}+\Delta]$ are in $U_{k_0}$. So 	
		\begin{align}
		E^{\star}-E^{\dagger}&=E_\text{t}(\mathbf{t}^{\star},\mathbf{p}^{\star})+E_{\text{s}}^{(b)}(\mathbf{x}^{\star})+\beta E_{\text{s}}^{(u)}(\mathbf{x}^{\star}) - E_\text{t}(\mathbf{t}^{\dagger},\mathbf{p}^{\dagger})-E_{\text{s}}^{(b)}(\mathbf{x}^{\dagger})-\beta E_{\text{s}}^{(u)}(\mathbf{x}^{\dagger}), \notag \\
		&{\geq} E_\text{t}(\mathbf{t}^{\star},\mathbf{p}^{\star})-E_\text{t}(\mathbf{t}^{\dagger},\mathbf{p}^{\dagger}) , \notag\\
		&\geq 0
		\end{align} Similar to Case 1, we can prove the inequality holds.
\end{case}
\section*{Appendix C: Proof of Lemma 3}
From (\ref{inequality between t and p}), it is clear that setting
\begin{align}
p_v=\max_{k\in \mathcal{K}} \left\{ \frac{n_0}{h_k^2} \left(2^{\frac{y_{k,v}RT}{Bt_v}}-1\right)\right\},\quad v \in \overline{\mathcal{V}}
\end{align}
will not loss optimality. Thus, we have:
\begin{align}
E_\text{t}(\mathbf{t},\mathbf{p})=\sum_{v\in \overline{\mathcal{V}}} t_v p_v=\sum_{v\in \overline{\mathcal{V}}} t_v \max_{k\in \mathcal{K}} \left\{ \frac{n_0}{h_k^2} \left(2^{\frac{y_{k,v}RT}{Bt_v}}-1\right)\right\}=E_1(\mathbf{t},\mathbf{p})
\end{align}
and the constraint in (8) can be eliminated. From lemma~2, we have:
\begin{equation}
E_{\text{s}}^{(b)}(\mathbf{x})=\sum_{v\in \overline{\mathcal{V}} \setminus \mathcal{V}}x_v E_b=E_b\sum_{v\in \overline{\mathcal{V}} \setminus \mathcal{V}}\max_{k\in \mathcal{K}} y_{k,v}=E_2(\mathbf{y})
\end{equation}
By (4), we have $1-y_{k,r_k}=\sum_{v\in \overline{\mathcal{V}}_{r_k}^+ } y_{k,v}$. Thus, we have:
\begin{equation}
E_{\text{s}}^{(u)}(\mathbf{y})=\sum_{k\in \mathcal{K}} (1-y_{k,r_k})E_{\text{u},k}=\sum_{k \in \mathcal{K}} E_{u,k} \sum_{v\in \overline{\mathcal{V}}_{r_k}^+}y_{k,v}=E_3(\mathbf{y})
\end{equation}
Therefore, we can equivalently transform Problem~\ref{View selection and resource allocation} to Problem~\ref{view selection and time allocation}.
\bibliographystyle{IEEEtran}

\end{document}